\documentclass{article}

\usepackage{amsmath}
\usepackage{mathrsfs}
\usepackage{amsthm}
\usepackage{cite}
\usepackage{graphicx} 
\usepackage{subfig} 

\usepackage{times,amssymb,mathptm,color,epsfig}
\usepackage{fancybox,picinpar,wrapfig,tabularx,fancyhdr}
\begin{document}
\bibliographystyle{comnet}

\title{A new framework for dynamical models on multiplex networks.}


\author{Daryl R. DeFord and Scott D. Pauls}

\maketitle

\newtheorem{thm}{Theorem}
\newtheorem{Remark}{Remark}
\newtheorem{proposition}{Proposition}
\newtheorem{corollary}{Corollary}

\begin{abstract}
{Many complex systems have natural representations as multi-layer networks. While these formulations retain more information than standard single-layer network models, there is not yet a fully developed theory for computing network metrics and statistics on these objects. We introduce a family of models of multiplex processes motivated by dynamical applications and investigate the properties of their spectra both theoretically and computationally.  We study special cases of multiplex diffusion and Markov dynamics, using the spectral results to compute their rates of convergence.  We use our framework to define a version of multiplex eigenvector centrality, which generalizes some existing notions in the literature.   Last, we compare our operator to structurally-derived models on synthetic and real-world networks, helping delineate the contexts in which the different frameworks are appropriate. }

\end{abstract}
\newpage

\section{Introduction}
Multi-layer networks form useful generalizations of single-layer networks, providing annotation to sets of edges in a network.  As we observe complex systems in detail, multi-layer data arises easily and naturally --- for example in social networks with different labels on interactions or trade networks with different categories of exchange.  We've seen an explosion of work exploring and analyzing multi-layer networks (see, for example, \cite{kivela_multilayer_2014,Boccaletti2014,salehi2015,physicsmulti}).  Much of this work focuses on a basic questions: how do we most effectively mathematically analyze a multi-layer network?  At two extremes, we can directly apply methods from single layer network analysis.  On one hand, we can view each layer as an isolated network, the collection of which we call the {\em disjoint layers model}, while on the other, we can aggregate all the data from the layers to form a single layer network.  Both of these methods are sometimes useful but are not always well adapted to multi-layer networks as they each ignore or conflate some of the aspects of the multi-layer presentation.  In between these poles, there are numerous ways to incorporate all of this information into a single model.

Different authors approach the transformation of multi-layer data into a multi-layer network in different ways.  Depending on the questions they are interested in, some authors take  structural approaches, which are analogous in different ways to the representation of single-layer networks using adjacency matrices, while others choose to instead model dynamic processes.  One core issue is that, in contrast to the single-layer case, structural and dynamic models for multi-layer networks do not have obvious linkages, which limits the overall usefulness of either type of model.  In particular, analogues of dynamic operators derived from structural representations, such as graph Laplacians and Markov chains for single-layer networks, are inappropriate models for some types of dynamics on multi-layer networks.

We have two goals in this paper.  First, we focus on examples of multiplex networks --- an important subset of multi-layer networks --- where treating copies of the same node in different layers as separate entities is inappropriate.  This class of networks includes many important examples including social networks.  Second, we formulate our models  through the lens of dynamics with the goal of representing these processes as accurately as possible.  The resulting family of models can represent a wide variety of possible dynamics and, in special cases, recover some existing work such as an information flow and consensus model  \cite{trpevski_discrete-time_2014}, some of the random walk models of \cite{domenico_navigability_2014}, as well as dynamic models used to define one class of centralities on multiplex networks \cite{sola_eigenvector_2013}.

For the types of applications we restrict to, existing structural models introduce distortions in associated dynamic models. One of our main goals is to avoid a common structural feature in multiplex models: interactions between multiple copies of the nodes associated to the different layers.  In instances where these copies are a mathematical convenience rather than an aspect of the modeled system, connections between node copies can introduce confounding feratures into dynamic processes.  We carefully construct our models to avoid these types of distortions by crafting inter-layer dynamics that do not involve exchanges between copies of nodes on different layers but instead permits nodes on one layer to interact directly with those on other layers by allowing effects to pass through from one layer to another.  Throughout this work, we will compare our results not only to the aggregate and disjoint layers models described above, but also to an existing structural model, the supra-adjacency matrix and an associated dynamic operator, the supra-Laplacian \cite{gomez_diffusion_2013}.  This popular model is best adapted to the case we avoid --- when inter-layer connections model actual features of the system.  Consequently, the comparisons help delineate the impact of the different modeling choices.  

One of the themes we revisit throughout this work is the idea that different versions of our dynamic operator interpolate between the two extremes described above --- dynamics on the averaged aggregate network and on the disjoint layers model.  At one extreme, the simplest version of our model, which we call the {\em equi-distribution model}, has a great deal in common with the averaged aggregate network, while the most general version of our model, the {\em general mixing model}, generically has some properties that approach those of the disjoint layers.  This flexibility is a strength of our framework as even the most general forms of other dynamic models in the literature fall in a more restricted range of this interpolation.

Connections between instances of our operator and different aggregations of the data form a second theme in this paper.  We see that three special cases of our general mixing model --- the equi-distribution model described above, a more general version we call the {\em ranked layers model}, and the still more general {\em unified node model} --- have spectral properties in common with specific weighted aggregations of the layer networks.  Further, if we enforce dynamics that treat all the instances of a node across the layers as a single entity, any multiplex model necessarily aggregates to {\em some} single-layer network.  For our models, we find that the resulting aggregations reflect the intra-layer dynamics appropriately weighted by aspects of the inter-layer interactions. 

As methods for data collection become more detailed, faster, and more complete, we expect that data from complex systems will naturally have a large number of layers:  finer granularity in observation allows us to make finer distinctions.  The more layers we have associated to a fixed system, the sparser we expect them to be --- the total description of the system is spread thinner and thinner.  Recently collected social network data from $75$ villages in the Karnakata region of India demonstrate the beginning of this trend \cite{Banerjee2013}.  Researchers collected data concerning $12$ types of interaction, yielding $12$ layers, all of which are quite sparse.  This type of study is not unique -- a similar study was recently carried out in Honduras tracking the spread of public health information and interventions \cite{Kim2015} -- and represents a new paradigm in social science research.  Another strength of our model is that it scales well with the number of layers.

We explore the properties of our models in three dynamical settings --- diffusion, random walks, and an analogue of eigenvector centrality.  In each of these cases we see examples of the broad results discussed above.  In examining diffusion, we see a connection between the spectrum of our multiplex Laplacian operator and that of sums of reweighted graph Laplacians of the layers in our simplified models.  This provides an avenue to understand the rate of convergence of the diffusion process, where we find examples of super-diffusion --- when the rate for the multiplex operator exceeds the maximum of the rates of the individual layers --- that depend subtly on both the inter-layer dynamics as well as the topologies of the layers.  The supra-Laplacian \cite{gomez_diffusion_2013,sole-ribalta_spectral_2013} exhibits super-diffusion in some instances when the inter-layer diffusivity constant is large.  In our case, the rate peaks not due to ramping up the inter-layer transfer, but due to a particular regime of relative relevances among the layers, revealing a new mechanism through which diffusion in multiplex networks is greater than the sum of its parts.

We find similar connections between the steady-state vectors for our stochastic dynamical operators and sums of reweighted stochastic operators associated to the layers of the multiplex network, and connections to the rate of convergence of the random walk process.  Here we see the clearest evidence of what happens as the number of layers grows.  For our operators, the rate of convergence remains essentially constant as we increase the number of layers.  In contrast, stochastic processes associated to the aggregate network converge faster and faster with the number of layers, while those associated to the supra-adjacency matrix (the structural matrix underlying the supra-Laplacian) converge more and more slowly.  This last fact is a consequence of the choice of using interacting copies of nodes in the dynamic model --- as the number of layers grows, each group of node copies forms a complete clique in which the resulting random walk process become more and more likely to linger.

We derive eigenvector centrality analogously to the derivation in single-layer networks, finding the clearest evidence that our family of operators interpolates between the aggregate and disjoint layer models.  Eigenvector centralities of the general mixing model are closest to those of the disjoint layers model while those of the equi-distribution model closest to the aggregate, with the others in between.  We note that using the supra-adjacency matrix to calculate centralities completes one endpoint of this interpolation as it yields centrality scores which are essentially linear transformations of those of the disjoint layers.

While this combination of theoretical and simulation results provides a rich description of the properties of our family of operators,  we augment them with analysis of multiplex networks arising from two sets of empirical data, the Karnakata village social networks and the World Trade Web (WTW).  We find in both cases that the results for these empirical multiplex networks are consistent with our earlier work and, in particular, that using synthetic network models provide reasonable predictions about the behavior for these examples.  For the World Trade Web, we measure random walk betweenness centrality for the unified node model and compare it to that of the aggregate and disjoint layers models.  We find that the unified node model, as an interpolation between the two, reveals impacts of trade asymmetries that the other models can miss.  In particular, petroleum-rich countries, where fuel exports dwarf other trade, show how the interaction between the different layers impacts the overall centrality for the unified node model appropriately.

For the Karnakata social networks, we consider a scenario where individuals are learning medical information via interactions within their social network.  In this case, we model one layer --- the ties identifying trust on medical issues --- having more importance than the others using the ranked layers model.  We find that these empirical examples behave in line with our theoretical and simulation results and exhibit super-diffusion of information flow in some cases.

We organize the balance of the paper as follows.  In Section \ref{sec:motivation} we further elaborate on the motivation for this work, and introduce our two empirical examples.  In Section \ref{sec:framework} introduces the mathematical modeling framework.  Section \ref{sec:spec} examines spectral properties of our family of models from using both theoretical investigations and simulation.  We also interpret these results in the context of  three dynamic processes.  Finally, in Section \ref{sec:apps} we use our framework to examine the World Trade Web and the Karnakata village social networks.

\section{Motivation}\label{sec:motivation}

Even the initial steps in representing data as a multiplex network leads to problems with the interplay between structure and dynamics. Suppose we have a multiplex network with node set $N$ of size $n$ and a set of $k$ layers with edge sets $E^\alpha, \alpha \in {1,\dots k}$. To layer $\alpha$ we associate an adjacency matrix $A^\alpha$ to $E^\alpha$ where $A^\alpha_{ij}$ is the weight of the edge between node $j$ and $i$ for zero if there is no edge between those nodes in layer $\alpha$.  While there are various ways we can wrap this information up --- we could use one of our two poles, considering the averaged aggregate network $\frac{1}{k}\sum_\alpha A^\alpha$ or the disjoint layers, $\{A^\alpha\}$, or combine them other ways that leave the layers distinct but allow them to interact ---  every choice is a compromise losing or conflating aspects of the multiplex data.  However, the two poles form guideposts in the analysis of multiplex network models --- comparing a new model to the aggregate and disjoint layers models helps to illuminate its properties in context.

Even when opting for the method with the seemingly most flexible approach where we model interactions between the layers, there is an inherent tacit assumption: each node exists as a collection of copies of itself, one for each layer.  Many of the models in the literature are of this form --- for example, the tensorial and supra-adjacency multiplex formulations \cite{cozzo_structure_2013,gomez_diffusion_2013,domenico_navigability_2014}.  We will call a structural model of this type a {\em matched sum}, where the multiplex model is a direct sum of the layer networks with the node copies on each layer matched to one another and coupled together via inter-layer connections.  We note that researchers studying temporally evolving networks using a multi-layer structure (e.g. \cite{mucha_community_2010}), use a variant of this idea, where only the nodes of temporally adjacent layers are connected to one another.  For simplicity, we will not consider this type of sum in our work.

The supra-adjacency formulation, for example, encodes the entire multiplex network in an $nk \times nk$ block matrix which the layer adjacency matrices along the diagonal and the off diagonal blocks given cross-layer connectivity.  Most often, the diagonal blocks are given by the adjacency matrix of the layers, while the off-diagonal blocks are copies of the identity --- the copies of the nodes are linked to one another:
\[\mathfrak{M}=\begin{pmatrix} A^1 &  \dots & I\\ \vdots & \ddots & \vdots \\ I & \dots & A^k\end{pmatrix}.\]
While the matched sum is appropriate in instances where the multiplex network represents a larger annotated single-layer network in which the node copies have distinct properties, it can introduce difficulties in cases where the nodes are indivisible in some sense.  In the latter case, dynamical processes defined in terms of multiple node copies can create confounding effects.  In our approach to the this case, we focus on modeling dynamic processes as accurately as possible, ignoring (for now) structural representations.  We do this for two reasons.  First, many of the questions researchers pose about networks --- how fast does information or disease spread in a population?  what is the maximum flow between two nodes in a network?  how does a random walk evolve? --- are naturally dynamic.  Second, unlike for single-layer network, as hinted at above, there is no natural link between structural representations and natural dynamical processes on multiplex networks.  While we can often easily describe intra-layer dynamics via analogues of single-layer network models, deciding on appropriate inter-layer dynamics is a more difficult task.

To help demonstrate further motivations for our approach, we describe two examples to which we return throughout the paper.  These examples demonstrate two instances where, for different reasons, treating node copies separately is problematic and serve as concrete demonstrations of situations where our framework can be useful.  

\subsection{Information flow in social systems}\label{sec:introsocial}
 Social systems are naturally multiplex or multi-layer --- individuals often have multiple partially overlapping social arenas which they use in different ways for different purposes.  Recently, researchers collected a complete snapshot of the overlayed social systems in 75 small Indian villages \cite{Banerjee2013}.  The interviewers asked each member of the village to identify different aspects of their social networks via specific markers --- kin identification, shared religious observance, advice seeking, food loaning or borrowing, seeking medical help, etc. --- painting a picture of a rich multiplex social tapestry.  They used these networks to understand the spread of information --- in this case, how information about a micro-loan program percolated through the communities. In \cite{Banerjee2013}, the authors aggregated the layer data into a single-layer network and analyzed information flow throughout.  We will use the full multiplex data, where we represent each layer as a binary adjacency matrix associated to the survey data.

In such a network, we can model intra-layer dynamics using any standard information flow dynamics --- diffusion, epidemic models, etc. --- modified to reflect the nature of the network.  Inter-layer dynamics must model how information passes through nodes --- the individuals in a community --- and is potentially transferred to other individuals across layers.  Such dynamics can be complex as they depend substantially on how the individual node views the utility of the different layers in which they participate. Additionally, many data sets of this type contain one or more distinguished layers relevant to a particular topic, such as medical advice relations in the context of public health interventions. The dynamical effects of these layers can be obscured by either partial or complete aggregation.

We see a simple mechanism for information flow in this type of system.  Individuals collect pieces of information from their contacts in different layers of the network. Aggregating the informational components, they then redistribute the information based on their perception of the different layers with respect to the information transferred.  This two step interaction between the intra- and inter-layer dynamics presents new challenges in describing the overall dynamics.  The dynamics track information propagation and, at any given time, the nodes possess fixed pieces of that information.  While the nodes then choose how to further disseminate the information, they still possess the entirety of the information at that time. 

 \subsection{International trade}\label{sec:introWTW}  Trade is the core of economic interactions and plays an important role in illuminating aspects of international action between state actors.  Bilateral trade between states reveals aspects of facilitated cooperation between states, even when they are at odds politically.  As such, trade serves as a marker for economic interdependence which interacts with political interdependence as well as political action, revealing connections with conflict, alliances, and economic risk \cite{Moaz2009,Long2006,Li2010,foti_stability_2013}.

States report trade data yearly, recording import and export totals for the various goods they trade with their partners.  Given a list of commodity types labeled $\{1,\dots,k\}$ traded between $n$ states, this data naturally yields a multiplex network:  each layer is a weighted directed network where $A_{ij}^\alpha$ has weight equal to the dollars paid from $j$ to $i$ for importing good $\alpha$.  For each layer, this information defines a dynamic process, giving the flow of money (or dually, goods) between different states.  Using these descriptions of the intra-layer dynamics, we are left to decide how to encode the inter-layer dynamics, which rest on determining how goods and money transform and circulate within the given node before leaving as an exported good or as payment for imports.

There are two ways dynamics on individual layers can transfer between layers.  Within a given country, the money gained from exports circulates in exchange for other goods and services, some of which might be imported from other countries.  Similarly, people either use imported goods internally or as components in the production of other goods and services, some of which may be then exported in the future.  While there is data on aspects of this process, for example in input-output tables \cite{L1986}, much of it is inaccessible due to the difficulties of accurate and timely data collection.  Another more serious limitation stems from the rate of data collection:  these processes generally take place on a different time scale --- the internal economic churning primarily happens within the year between import/export reports.

Given the data restrictions, we can define a simplified dynamics.  From the vantage point of a single state, we see money from exports flow into the country, exchange internally, and then, in part, leave the country as payment for imported goods.  Viewing this in three steps, we first have money flowing in, then aggregating within the country, and last, a portion flowing out to other countries.

 \section{Mathematical Framework}\label{sec:framework}
Choices in constructing multiplex models are broadly ones of encoding heterogeneity --- differences in how nodes treat the quantities transferred and the layers themselves, differences in time scales, and differences of interpretation between intra- and inter-layer dynamics. To fix notation, we assume that there are $n$ nodes in the multiplex, labeled $\{1,\dots, n\}$ and $k$ layers, $\{1,\dots k\}$.  Generally, we will use the convention that subscript roman letters (e.g. $i$ and $j$) to refer to nodes and superscript Greek letters (e.g. $\alpha$ and $\beta$) to refer to layers.  Like other treatments of multiplex and multi-layer networks, as a notational convenience we will initially use $k$ copies of each node, one for each layer to begin to encode the dynamics, which we denote $n^\alpha_i$.  This violates our principle of {\em not} treating a single node as a collection of copies, but as we construct the dynamics, we'll be careful to not let these copies interact in the end.  On each layer $\alpha$, we have an edge set $E^\alpha$.

To model the intra-layer dynamics, first let $D^\alpha$ denote the dynamic operator on layer $\alpha$ and combine all the layer operators together into a diagonal block matrix:
\[D = \begin{pmatrix}D^1 &  \dots & 0 \\  \vdots  & \ddots & \vdots\\ 0 & \dots  &D^k\end{pmatrix}.\]

\begin{figure}
\centering
\includegraphics[height=2in]{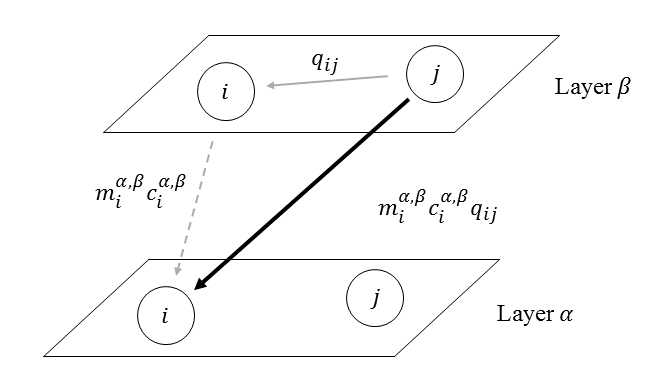}
\caption{A schematic of our proposed dynamics between $n^\beta_j$ and $n^\alpha_i$.  The two gray arrows show how the overall flow (black arrow) breaks into components --- the flow on layer $\beta$ (solid gray arrow) and the transfer to layer $\alpha$ (dotted gray arrow). }\label{fig:basic_dynamics}
\end{figure}

Next, we can represent the portion of the quantity moving through the multiplex as an $nk \times 1$ vector $v=(v^1,\dots,v^k)^T$ where $v^\alpha$ gives the quantities moving in layer $\alpha$ at that specific time.  Throughout the paper, we use the convention that for an $nk \times 1$ vector $w$, $w^\alpha_i$ is the $(i+k(\alpha-1))$ component of $w$ --- i.e. the quantity at node $i$ in layer $\alpha$.   Consequently, $Dv$ represents the effects of the intra-layer dynamics.  To define our inter-layer dynamics and how the two processes interact, consider $n^\beta_j$ and one of its neighbors $n^\beta_i$.  If some quantity passes from $n^\beta_j$ to $n^\beta_i$ within layer $\beta$ according to the intra-layer dynamics, we then allow a portion of that quantity to pass to $n^\alpha_i$, perhaps dilated by external factors, in the same time step.  We define $c^{\alpha,\beta}_i$ to be the proportion of the quantity at node $i$ on layer $\beta$ that passes through to layer $\alpha$.  By this convention, we have that $\sum_{\alpha =1}^k c^{\alpha,\beta}_i=1$ for all $\beta \in \{1,\dots,k\}$ and $i\in \{1,\dots n\}$. To model external dilations, we let $m^{\alpha,\beta}_i$ be a multiplier on node $i$ from layer $\beta$ to layer $\alpha$. Including the $m^{\alpha,\beta}_i$ allows us to scale the $c^{\alpha,\beta}_i$ to achieve any linear combination instead of the proportional split enforced by $\sum_{\alpha =1}^k c^{\alpha,\beta}_i=1$. We view $m^{\alpha,\beta}_i$ as a reflection of how much $n^\alpha_i$ values quantities flowing in from layer $\beta$. A schematic of this two-step description is given in Figure \ref{fig:basic_dynamics}, demonstrating how direct flows between node copies are removed from the dynamics by incorporating them into the overall flow between $n^\beta_j$ and $n^\alpha_i$.

To summarize, we update the vector $v$ to a new vector $v'$ according to the formula:
\begin{equation}\label{eqn:full_dynamics}(v')^\alpha_i =\sum_{\beta=1}^k m^{\alpha,\beta}_i c^{\alpha,\beta}_i (Dv)^\beta_i.
\end{equation}
To present this in linear algebraic language, we let $C^{\alpha,\beta}$ be the diagonal matrix with diagonal given by $(m^{\alpha,\beta}_1 c^{\alpha,\beta}_1, \dots ,m^{\alpha,\beta}_n c^{\alpha,\beta}_n)$ and $C$ be a block matrix with the $(\alpha,\beta)^\text{th}$ block given by $C^{\alpha,\beta}$.  Then, we can summarize \eqref{eqn:full_dynamics} as
\begin{equation}\label{eqn:la}
v'=\mathfrak{D} v = CDv=\begin{pmatrix} C^{1,1} & \dots &C^{1,k} \\ \vdots& \ddots & \vdots \\ C^{k,1} & \dots & C^{k,k} \end{pmatrix} Dv =
\begin{pmatrix} C^{1,1}D^1 & \dots &C^{1,k}D^k \\ \vdots& \ddots & \vdots \\ C^{k,1}D^1 & \dots & C^{k,k}D^k \end{pmatrix}v
\end{equation}

We note that if $C^{\alpha,\alpha}=I$ and $C^{\alpha,\beta}=0$ for $\alpha \neq \beta$, we recover the disjoint layers model which is simply given by $D$ where $D^\alpha$ is the adjacency matrix, $A^\alpha$ of the layer. As we will see below, a simpler version of this general operator is closely linked to the average aggregate operator, $\frac{1}{k}\sum_{\alpha=1}^k A^\alpha$ when $D^\alpha=A^\alpha$.  This reflects one of our recurring themes, that our operator interpolates between these two simple models.

If all of the $m^{\alpha,\beta}_i=1$, we say that $\mathfrak{D}$ is {\em closed}.  Consequently, when $\mathfrak{D}$ is closed, $C$ is stochastic, which we will use when discussing random walk dynamics on multiplex networks in Section \ref{sec:rw}.  When $\mathfrak{D}$ is closed and the $D^\alpha$ are random
walk matrices we recover models used in \cite{domenico_navigability_2014} and \cite{trpevski_discrete-time_2014}.

The use of the matrix $C$ places the intra-node activity on a different time scale than that of the inter-layer dynamics. Thus, our framework is particularly applicable in cases, like those outlined in Section \ref{sec:motivation}, where the inter-layer dynamics have much shorter time scales than the intra-layer dynamics, or when the intra-node dynamics are obscured or absent.  However, if intra-node dynamics are explicit and on a similar time scale to the inter-layer dynamics, dynamics built on a structural model such as the supra-adjacency matrix \cite{gomez_diffusion_2013} may be a more appropriate formulation.

Almost completely unconstrained choices for the $m^{\alpha,\beta}_i$ and the $c^{\alpha,\beta}_i$ make the framework extremely flexible, allowing us to encode very heterogeneous cross-layer dynamics,  which we refer to as the {\em general mixing} model.  Empirically determining all the coefficients in such a model requires an enormous amount of detailed data, which is often unavailable in practice.  Motivated by our trade and social network examples above, and the constraints enforced by their respective data sets, we discuss three simplifications related to the issues they raise.
\subsection{Unified nodes}\label{sec:uninodes}
Both our empirical networks in Section \ref{sec:motivation} present cases where the nodes first collect together the quantities before distributing the aggregate amount among the layers in the next time step.  Social networks present this most clearly, where individuals assimilate all of the pieces of information from their various sources before communicating them again.  In the situation we described for trade networks, where the data does not described the intra-node dynamics, the data constraints yield the same type of situation -- at the end of each step we collect the dollars within each node before beginning the trading anew.

Within our general framework, this type of aggregation imposes new constraints on the $\{c^{\alpha,\beta}_i\}$.  After applying the intra-layer dynamics, we aggregate all of the quantities in different layers each node $i$,
\[q(i)= \sum_{\beta =1}^k (Dv)^\beta_i.\]
Note that this is simply the projection of all of the layer quantities onto the nodes.  We then redistribute the aggregate quantity among the layers according to fixed proportions, $c^\alpha_i$, and externalities, $m^{\alpha}_i$,
\[(v')^\alpha_i = m^{\alpha}_i c^\alpha_i q(i)=m^{\alpha}_i c^\alpha_i \sum_{\beta =1}^k (Dv)^\beta_i,\]
where $\sum_{\alpha =1}^k c^\alpha_i=1$.  Comparing this to Equation \eqref{eqn:full_dynamics}, we see that these dynamics are a simplification of the full dynamics where $c^{\alpha,\beta}_i = c^\alpha_i$ and $m^{\alpha,\beta}_i=m^\alpha_i$ for all $\beta$.  In these situations, $C^{\alpha,\beta}\equiv C^\alpha = diag (m^{\alpha}_1 c^{\alpha}_1, \dots ,m^{\alpha}_n c^{\alpha}_n)$ and
\begin{equation}\label{eqn:la_unified}
v'=\mathfrak{D} v = CDv=\begin{pmatrix} C^{1} & \dots &C^{1} \\ \vdots& \ddots & \vdots \\ C^{k} & \dots & C^{k} \end{pmatrix} Dv =
\begin{pmatrix} C^{1}D^1 & \dots &C^{1}D^k \\ \vdots& \ddots & \vdots \\ C^{k}D^1 & \dots & C^{k}D^k \end{pmatrix}v
\end{equation}

For the WTW, given the opaqueness of the internal rearrangement we do not have fine enough detail in the available data to estimate parameters for the general mixing model, but we have sufficient information to use the unified node simplification.  To define the $c^\alpha_i$, we use a simple, general method to measure the importance of layer $\alpha$ to node $i$.  We use the in-degree of node $i$ in layer $\alpha$ as a proxy for this importance and normalize it by the aggregate in-degree:
\begin{equation}\label{WTW_params}
c_i^\alpha = \frac{\sum_{j} A^\alpha_{ij}}{\sum_{j,\beta}A^\beta_{ij}}.
\end{equation}
For our application below considering random walks on the WTW, we also require the model be closed.
\subsection{Ranked layers}\label{sec:hierlay}
In social systems, individuals place different values on the information coming from different layers depending on their relevance to the desired outcome.  For example, individuals seeking medical information would likely value the layer of people they identify as knowledgeable about medical issues over layers oriented towards other interactions, such as borrowing money.  If such distinctions extend over the entire set of nodes, we can model this using a ranking of layer importance that holds for all nodes.

The simplest version of this idea is a reduction of the unified node model which we call the {\em ranked layers model}, where we enforce the new conditions that $c^\alpha_i=c^\alpha_j$ and $m^{\alpha}_i=m^{\alpha}_j$ for all $i,j \in \{1,\dots,n\}$. In this case, $C^{\alpha,\beta}$ is a linear homothety given by $m^{\alpha}c^\alpha I$ and
\begin{equation}
v'=\mathfrak{D}_h v = CDv=
\begin{pmatrix} m^{1}c^1D^1 & \dots &m^{1}c^1D^k \\ \vdots& \ddots & \vdots \\ m^{k} c^kD^1 & \dots & m^{k}c^kD^k \end{pmatrix}v
\end{equation}\label{eqn:la_hier}

While the unified node model is one generalization of this, we can form a second model, the {\em generalized ranked layers model}, by allowing the constants to depend on both the source and target layers:  letting $c^{\alpha,\beta}_i=c^{\alpha,\beta}_j\equiv c^{\alpha,\beta}$ and $m^{\alpha,\beta}_i=m^{\alpha,\beta}_j\equiv m^{\alpha,\beta}$ for all $i,j \in \{1, \dots n\}$.  Now, $C^{\alpha,\beta}$ is a linear homothety given by $m^{\alpha,\beta} c^{\alpha,\beta} I$.  This is the asymmetric influence matrix $W$ introduced in \cite{sola_eigenvector_2013} for measuring eigenvector centrality in multiplex networks.   When the $D^\alpha$ are the adjacency matrices for the layers, we recover their computation of the global heterogeneous eigenvector centrality.  In both the ranked layer and generalized ranked layer models, $C$ can be written as the tensor of a $k\times k$ matrix $\tilde{C}$ and the $n\times n$ identity matrix.  Consequently, $\tilde{C}$ is a ``network of layers''  as defined in \cite{sole-ribalta_spectral_2013}. 

Our example social multiplex network from the Karnakata Village data provides a good example of where the generalized ranked layers model works well. In Section \ref{sec:social}, we'll consider exactly the situation described above where we model the spread of medical information through this multiplex network.  Using the generalized ranked layers model to privilege flows from the medical layer over others models our assumption that individuals trust medical knowledge transmitted through this layer more than others.   Like the WTW example, the data is not sufficient for anything more granular than this model:  the data collection did not include measures of trust at all, and certainly not on the level of the individual.  To encode this, we use a parameter $w$ to adjust the weight of the medical layer relative to the $11$ other layers given in the data, letting $c^{\textrm{*, medical}}=w$ and $c^{\textrm{*,non--medical}}=\dfrac{1-w}{11}$, allowing $w$ to vary between $\frac{1}{12}$ and $1$. We further assume the dynamic operator is closed, so that $m^{\alpha}=1$.

\subsection{Equi-distribution}\label{sec:equi}
Our last example is the most simplified, where we assume that all of the inter-layer distributions are equal, in other words that $c^{\alpha,\beta}_i=\frac{1}{k}$ and $m^{\alpha,\beta}_i=1$ for all $\alpha,\beta$ and $i$.  Consequently,
\begin{equation}
v'=\mathfrak{D}_e v = CDv=\frac{1}{k}\begin{pmatrix} D^1 & \dots &D^k \\ \vdots& \ddots & \vdots \\ D^1 & \dots & D^k \end{pmatrix}v.
\end{equation}\label{eqn:la_equi}
This model is a special case of the ranked layers model --- note that using the parameter $w=\frac{1}{12}$ in the last example reduces to the equi-distribution model.

In practice, this case arises when little is known about the inter-layer dynamics, or when the data is not rich enough to describe more general dynamics, and we choose a parsimonious model.  More importantly, as we will see below, the equi-distribution model provides us with a close link to the averaged aggregate network of a multiplex system.
\section{The spectrum and its relation to dynamics}\label{sec:spec}
As the spectrum of dynamic operators yield a great deal of information about single-layer networks and their dynamics, we next explore the spectral properties of $\mathfrak{D}$ both generally and in the special cases delineated in the previous section.  We'll often use a comparison to our two basic models associated to multiplex data --- the average aggregation and the disjoint layers models --- as well as the matched sum model used in much of the literature to place our results in context.  We will see that results for the matched sum are usually very close to the disjoint layers model while different versions of $\mathfrak{D}$ interpolate between the aggregate and disjoint layers model.

Our first theoretical results connect the spectra of $\mathfrak{D_u},\mathfrak{D_h}$ and $\mathfrak{D_e}$ to the spectra of different aggregations of the $D^\alpha$.

\begin{proposition}\label{prop:spec}
We consider the models $\mathfrak{D}_u, \mathfrak{D}_h,$ and $\mathfrak{D}_e$ and assume that the $C^\alpha$ are invertible for $\mathfrak{D}_u$ and that the $c^\alpha \neq 0$ for $\mathfrak{D}_h$.  Then,
\begin{enumerate}
\item (Unified Node Model) Let $D_a=D^1C^1+\dots D^kC^k$ and $\{(\lambda_i,w_i)\}$ be its eigendata.  If $\lambda_i \neq 0$, $(\lambda_i,v_i)$ is an eigenvalue/eigenvector pair for $\mathfrak{D}_u$ where \[v_i=(C^1 w_i,\dots, C^k w_i)^T.\]
\item (Ranked Layers Model) Let $D_a=m^1c^1D^1+\dots +m^kc^kD^k$ and $\{(\lambda_i,w_i)\}$ be its eigendata.   If $\lambda_i \neq 0$,  $(\lambda_i,v_i)$ is an eigenvalue/eigenvector pair for $\mathfrak{D}_h$ where
    \[ v_i = \left ( m^1c^1 w_i,\dots, m^kc^k w_i\right )^T.\]
\item (Equi-distribution Model) Let $D_a=\frac{1}{k}(D^1+\dots + D^k)$ and $\{(\lambda_i,w_i)\}$ be its eigendata.  If $\lambda_i \neq 0$  $(\lambda_i,v_i)$ is an eigenvalue/eigenvector pair for $\mathfrak{D}_e$ where $v_i=\frac{1}{k}(w_i,w_i,\dots,w_i)^T$.

\end{enumerate}
\end{proposition}

\noindent
{\em Proof: }  This theorem outlines some of the simpler cases where we can compute the spectrum of $\mathfrak{D}$ abstractly.  The defining equation for the eigenvalues and eigenvectors of $\mathfrak{D}$ is
\begin{equation}\label{eqn:thm1}
\mathfrak{D}v=\begin{pmatrix} C^{1,1}D^1 & \dots &C^{1,k}D^k \\ \vdots& \ddots & \vdots \\ C^{k,1}D^1 & \dots & C^{k,k}D^k \end{pmatrix}\begin{pmatrix} v^1 \\ \vdots \\ v^k\end{pmatrix} = \lambda \begin{pmatrix} v^1 \\ \vdots \\ v^k\end{pmatrix}.
\end{equation}
While there is not much we can do with this in general, when the $C^{\alpha,\beta}$ are not too complicated we can connect these equations to the analogous equations for appropriate weighted aggregations of the layers.  For $\mathfrak{D}_u$, Equation \eqref{eqn:thm1} reduces to the following system:
\begin{equation*}
\begin{split}
 C^1\left(D^1v^1 + \dots +D^kv^k\right )&= \lambda v^1\\
\vdots & \;\;\\
  C^k\left(D^1v^1 + \dots +D^kv^k \right )&= \lambda v^k.
\end{split}
\end{equation*}
Multiplying the $\alpha^\text{th}$ equation by $(C^\alpha)^{-1}$ yields $\sum_{\beta=1}^l D^\beta v^\beta=\lambda (C^\alpha)^{-1} v^\alpha$.  As the left hand sides of these equations are now all equal, we have that $\lambda (C^\alpha)^{-1} v^\alpha = \lambda (C^\beta)^{-1} v^\beta$.  As long as $\lambda \neq 0$, there is a vector $w$ so that $(C^\alpha)^{-1} v^\alpha=w$ for all $\alpha$.  Replacing $v^\alpha$ with $C^\alpha w$ makes all of the equations identical,
\[(D^1C^1+\dots+D^kC^k)w=\lambda w,\]
and the result then follows.  The second and third cases are special cases of this one.
\qed

Although determining the spectral structure of sums of matrices in terms of the spectra of the summands is difficult, if the matrices are symmetric we can use standard results due to Weyl (see \cite{fulton_eigenvalues_2000,knutson_honeycomb_1999}) to obtain upper and lower bounds on the individual eigenvalues of the derived operators. Next, we'll look more closely at the properties of the spectrum and their interpretations for more specific types of dynamics.
\subsection{Diffusion Dynamics}\label{sec:diffusion}
Modeling diffusion on networks as a discretization of the continuous heat flow yields the graph Laplacian whose spectral structure has strong connections to important graph properties such as connectivity, communities, as well as the evolution of random walks \cite{brualdi_mutually_2011,chung_spectral_1996,newman_networks_2010}.

To extend a diffusion model to the multiplex setting using our framework, we first examine the case where heat can flow both within and between layers via inter-layer connections.  We let $v$ be and $nk \times 1$ vector represent the temperature at each node copy and define the change in the value  of $v^\alpha_i$ with respect to time to be proportional to the sum of the differences in temperature between each node $i$ and its neighbors:
\begin{equation}
\frac{dv^\alpha_i}{dt}=-K\sum_{\beta=1}^kc_i^{\alpha,\beta}\sum_{n^\beta_i\sim n^\beta_j}(v^\beta_i-v^\beta_j).
\end{equation}
Here $K$ is the diffusion constant and the $c^{\alpha,\beta}_i$ represent the proportion of the effect on layer $\beta$ that passes through to $n^{\alpha}_i$. Under our assumptions about the nature of cross-layer connectivity, this definition is exactly analogous to the single-layer network case.  Linear algebraically, we summarize this as,

\begin{equation}\label{eqn:lap1}
\frac{dv^\alpha_i}{dt}=-K\sum_{\beta=1}^kc_i^{\alpha,\beta}(L v)^\beta_i,
\end{equation}
where
\[L=\begin{pmatrix} L^1 & \dots & 0\\ \vdots & \ddots & \vdots\\ 0 &\dots & L^k \end{pmatrix},\]
and $L^\alpha$ is the graph Laplacian associated to layer $\alpha$. Taken together, this is $\mathfrak{D}v$ when $\mathfrak{D}$ is closed and the layer dynamics are given by the respective layer graph Laplacians.  With this definition, Proposition \ref{prop:spec} connects the spectrum of this model of diffusion dynamics to that of a graph Laplacian formed from particular instances of weighted aggregated networks.  

We compare this to two other models --- the supra-Laplacian, which allows inter-layer diffusion solely between node copies, and the disjoint layers model, which doesn't allow inter-layer diffusion at all.  The spectrum of the supra-Laplacian has a connection to the spectrum of the averaged aggregate network for multiplex networks so long as the inter-layer coupling is relatively strong \cite{sole-ribalta_spectral_2013}.  The spectrum of the Laplacian associated to the disjoint layers model, given by the block diagonal matrix of the graph Laplacians on the layers,  is simply the union of the spectra of the layer Laplacians.  Taken together, this collection of results demonstrates how different instances of $\mathfrak{D}$ move away from the aggregate model in ways that allow us to place different emphases on different layers according to our inter-layer coupling coefficients.

\subsubsection{Enforcing node uniformity}\label{sec:enforce}
 So far, our diffusion model implicitly assumes that each copy of each node can be assigned a separate amount of the diffused quantity. This is a reasonable assumption for applications such as some economic exchange networks or the transportation models considered in \cite{domenico_navigability_2014}. However, in examples such as our social multiplex where the node copies are only representing different interaction types associated to the same individual, each individual has some fixed amount of information regardless of what types of interactions they are performing.

In such a case, we think of each copy of the node as contributing a change in temperature to the whole node according to Equation \eqref{eqn:lap1}.  Then, the total change in the temperature for the entire node is
\[\frac{dv_i}{dt}=\sum_{\alpha =1}^k \frac{dv^\alpha_i}{dt}=K\sum_{\alpha,\beta=1}^kc_i^{\alpha,\beta}(L^\beta v)^\beta_i  =K\sum_{\beta=1}^k(L v)^\beta_i,
\]
as $\sum_{\alpha=1}^k c^{\alpha,\beta}_i=1$.  Consequently, under the assumption of a single temperature per node, our model reduces to the Laplacian of the aggregation of the adjacency matrices of the layers, $A^1+\dots+A^k$.  Another way to see this is to recast the operator $\mathfrak{D}$, where $D^\alpha$ is the graph Laplacian $L^\alpha$ on each layer, to act on an $n \times 1$ vector of node temperatures $T$.  To do this, we first duplicate the temperature $T(i)$ to all the copies of $i$, apply $\mathfrak{D}$, and then sum up the results to get the change in temperature for the whole node.  Mathematically, we can realize this as follows:
\begin{equation*}
\begin{split} \frac{dT}{dt} & = \begin{pmatrix} I &\dots& I \end{pmatrix}  \begin{pmatrix} C^{1,1}L^1 & \dots &C^{1,k}L^k \\ \vdots& \ddots & \vdots \\ C^{k,1}L^1 & \dots & C^{k,k}L^k \end{pmatrix} \begin{pmatrix} I\\\vdots\\ I \end{pmatrix}T \\
&=  \begin{pmatrix} \sum_{\alpha =1}^k C^{\alpha,1} L^1&\dots& \sum_{\alpha=1}^k C^{\alpha,k} L^k \end{pmatrix} \begin{pmatrix} I\\\vdots\\ I \end{pmatrix}T \\
&= \begin{pmatrix} L^1 & \dots & L^k \end{pmatrix} \begin{pmatrix} I\\\vdots\\ I \end{pmatrix}T = \left (\sum_{\alpha=1}^k L^\alpha \right ) T.
\end{split}
\end{equation*}
In the second to last line, we've used that $\sum_{\alpha=1}^k C^{\alpha,\beta} = I$.

 In the case where $\mathfrak{D}$ is not closed, the $m^{\alpha,\beta}_i$ model external sources or sinks of heat in the layers, and the resulting total change in temperature for node $i$ is considerably more complicated:
\[ \sum_{\alpha =1}^k \frac{dv^\alpha_i}{dt}=K\sum_{\alpha,\beta=1}^km^{\alpha,\beta}_i c_i^{\alpha,\beta}(L T)^\beta_i.
\]

Requiring the nodes to have a single temperature necessarily requires us to aggregate the multiplex network into an appropriate single-layer network representation.  These calculations show that the simplest aggregation of the layers is the most appropriate when $\mathfrak{D}$ is closed --- the heterogeneities in the multiplex network do not impact the results. However, when $\mathfrak{D}$ is {\em not} closed, our framework provides a weighted aggregation that incorporates the heterogeneities across the layers, $\sum_{\alpha,\beta=1}^k C^{\alpha,\beta}L^\beta$ . We note that enforcing the same uniformity on the supra-Laplacian operator \cite{gomez_diffusion_2013} yields the aggregation $\sum_{\alpha=1}^k d_\alpha L^\alpha$ where the $d_\alpha$ are the intra-layer diffusion constants, irrespective of the inter-layer diffusion constants.  This reflects the different choice made in the construction of the supra-Laplacian model of diffusion --- as inter-layer diffusion happens only within nodes, if we force all node copies to have the same temperature effectively stops diffusion between the layers.   This contrast again points out how different modeling choices impact the results indicating the care one must take in choosing a model appropriate to the situation at hand.   

\subsubsection{Relation to the spectra of the layer Laplacians}\label{sec:speclayers}
The operators $\mathfrak{D}$ and $\mathfrak{D}_u$ generally aren't symmetric and consequently don't automatically share properties of the component layer graph Laplacians.  However, the equi-distribution and the ranked layers models are related to the individual layer Laplacians when the layers are undirected graphs. Using tools from the theory of Hermitian matrices we can prove the following bounds for the operator $\mathfrak{D}_e$ when the intra--layer dynamics are Laplacians.

\begin{proposition} Suppose all the layers of the multiplex network are undirected graphs. Then, if each  $D^\alpha$ is the graph Laplacian associated to layer $\alpha$ then $\mathfrak{D}_e$ and $\mathfrak{D}_h$ are positive semi--definite and eigenvectors corresponding to distinct non--zero eigenvalues are orthogonal.
\end{proposition}

 Consequently, the solutions to the differential equation $\frac{d\hat{\varphi}}{dt}+\mathfrak{D}\hat{\varphi}=0$ for these cases have the same algebraic and analytic structure as the standard Laplacian: the solution consists of constant elements and terms that decay exponentially. We now find eigenvalue bounds for $\mathfrak{D}_e$ in this case. We assume that the layer networks are connected --- although similar bounds exist in the case where the layer networks are disconnected, the formulas become more complex.

We introduce some additional notation. Let $\lambda^\alpha_\ell$ be the $\ell^{th}$ eigenvalue of $D^\alpha$ written in descending order with $\lambda_F^\alpha$ representing the Fiedler value \cite{F1,F2} corresponding to $D^\alpha$.  Let the eigenvalues of $\mathfrak{D}_e$ be $\lambda_1\geq\lambda_2\cdots\geq \lambda_{kn}$.  As $\operatorname{rank}(\mathfrak{D}_e)=n-1$, for $p>n-1$ we have $\lambda_p=0$.  Finally, let $m$ be the index such that $\lambda^m_1=\operatorname{max}_\alpha(\lambda_1^\alpha)$.
\begin{proposition}\label{prop:equilap}
 Suppose all the layers are connected and each  $D^\alpha$ is the layer graph Laplacian. Then, we have the following eigenvalue bounds for the operator $\mathfrak{D}_e$
\begin{enumerate}
\item Fiedler Value: $ \operatorname{max}_\alpha(\lambda_F^\alpha)\leq k\lambda_F \leq \lambda_F^m+\sum_{\beta\neq m} \lambda_1^\beta$,
\item Leading Value: $\operatorname{max}_i(\lambda_1^i)\leq k\lambda_1\leq \sum_{i}\lambda_1^i$,
\end{enumerate}

\end{proposition}
\noindent

These bounds are special cases of the following more general but less computationally feasible bounds:
$$\operatorname{max}_i(\lambda_{n-j}^i)\leq k\lambda_{n-j}\leq \min_{J\vdash n+k-(j+1)}\left(\min_{\sigma\in S_n}\left(\sum_{\alpha=1}^k \lambda_{j_\alpha}^{\sigma(\alpha)}\right)\right),$$
where $J=(j_1,j_2,\ldots,j_k)$ such that $\sum_{\alpha=1}^kj_\alpha=n+k-(j+1)$.
As remarked after Proposition 1, we can give a similar characterization of the eigenvalues of the ranked layers model, where we obtain results equivalent to Proposition 4.4, replacing $\lambda^\alpha_j$ with $(c^\alpha\lambda^\alpha)_j$ (after possible reordering) in each occurrence.
\subsubsection{The rate of convergence of multiplex diffusion}\label{sec:ratediffusion} While the theoretical results above set the stage, they provide a rather broad range of possible values.  We will refine these by looking at the Fiedler values of ensembles of synthetic networks. For single-layer networks, the second smallest eigenvalue of the graph Laplacian operator controls the convergence rate of the diffusive process. Here, we compare the bounds from Proposition \ref{prop:equilap} for the equi-distribution and ranked layers operators on simulated synthetic networks.

\begin{figure}[!h]
\centering
\subfloat[Equi-distribution]{{
\includegraphics[height=1.5in]{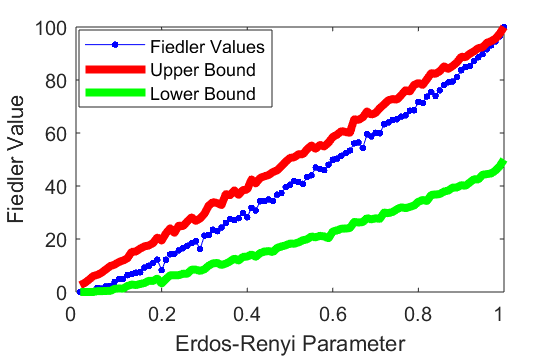}}}\qquad
\subfloat[Ranked Layers]{{
\includegraphics[height=1.5in]{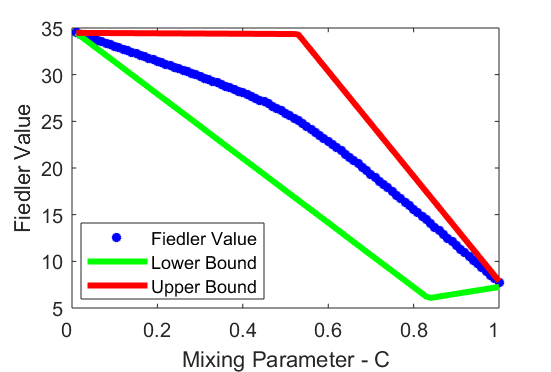}}}

\caption{(a) Comparison of Laplacian eigenvalue bounds for a two-layer equi-distribution model multiplex. (b) Comparison of Laplacian eigenvalue bounds for a collection of two-layer ranked layers model multiplex networks.}\label{fig:bounds}
\end{figure}

First, we construct a family of two-layer multiplex networks, where each layer is an Erd\H{o}s--R\'enyi graph with the connection probability varying from 0 to 1. For each of these parameter values we construct the equi-distribution operator and compute the associated Fiedler value as well as the corresponding bounds from Proposition \ref{prop:equilap}, with results shown in Figure \ref{fig:bounds} (a).  Our bounds are tight as we see them achieved at the extremes $p=0$ and $p=1$. In between, the observed Fiedler values tend to be closer to the upper bound, which we have observed in many examples, both with synthetic and empirical data.

Second, as in \cite{sole-ribalta_spectral_2013}, we analyze bounds for the ranked layers model using simulated two-layer networks.  With two layers, our construction has one parameter $c\in [0,1]$ which we assign to $c_i^{1,2}$ and $c_i^{1,1}$ for all $i$. By the further assumption that $\sum_\alpha c_i^{\alpha,\beta}=1$, we have $c_i^{2,1}=c_i^{2,2}=1-c_i^{1,2}$ for all $i$.  We construct two $100$--node Erd\H{o}s--R\'enyi layers with different densities, given by connection probabilities $p=0.2$ and $p=0.5$, and consider the corresponding multiplex operators where the intra--layer operators are the layer Laplacians and the mixing coefficients are given as above.

In Figure \ref{fig:bounds}b, we show the second smallest eigenvalues for these multiplex operators together with bounds analogous to those in Proposition \ref{prop:equilap} adjusted for by scaling the layer Fiedler values by $c$ and $1-c$.  The second eigenvalue smoothly varies between two extreme values as $c$ moves from zero to one.  These extremes are roughly equidistant from the bounds, but for $c$ near one half, where the ranked layers model reduces to the equi-distribution model, the eigenvalues are much closer to the upper bounds. Using other generative models for the layers yields similar results.

Next, we extend the analysis to a case of multiplex networks with more layers and varying layer topologies. Thinking ahead to our application to the Karnakata village social networks, we construct examples with similar features.  Each multiplex consists of 12 layers on 501 nodes. For each of the multiplex structures, eleven of the layers are constructed as Erd\H{o}s--R\'enyi graphs. For the twelfth, the first multiplex network has a Watts-Strogatz layer, the second a single path connecting all the nodes, and the third a layer consisting entirely of disjoint triangles. We then change the relevance of this layer via a parameter $w \in [\frac{1}{12},1]$ where $w=\frac{1}{12}$ yields the equi-distribution model and letting $w$ tend to one pushes the dynamics towards to the distinguished layer, as described in Section \ref{sec:hierlay}.

\begin{figure}
		\centering
\includegraphics[height=2in]{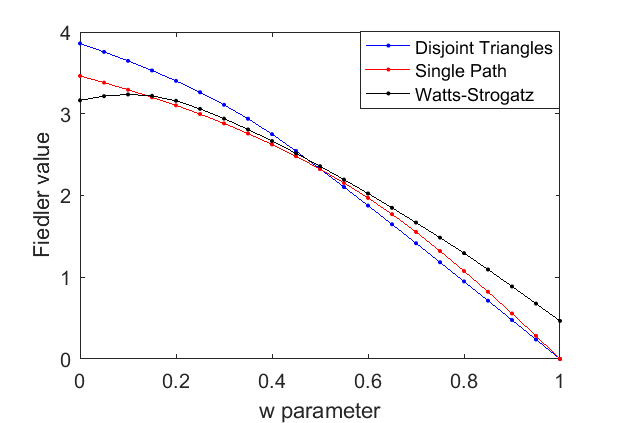}
	\caption{Diffusion results for three synthetic multiplex structures with distinguished layers. Each multiplex has eleven Erd\H{o}s--R\'enyi layers and one distinguished layer; Watts--Strogatz (black), a single path (red), and a collection of disjoint triangles (blue). The Erd\H{o}s--R\'enyi layers for the single path and disjoint triangle multiplex layers are denser than the Watts--Strogatz multiplex layer. However, as $w\rightarrow1$ the Watts--Strogatz multiplex has more rapid diffusion than the other two structures. }\label{fig:threems}
	\end{figure}

As we expect the diffusion on the first model to converge more quickly than that of the second and third as $w\rightarrow 1$, we construct the  Erd\H{o}s--R\'enyi layers of the latter two multiplex networks to be denser than the corresponding layers the first network. The results are displayed in Figure \ref{fig:threems}.	As expected, in the equi-distribution case, the second and third networks offer faster diffusion, as they correspond to denser networks. However, as $w$ increases, the first network overtakes the other diffusion rates, passing the path layer at roughly $w=0.3$ and the triangle layer at $w=0.55$. Thus, the rate of diffusion depends on both the weight selected as well as the topological structure of the distinguished layer.  This suggests that effective multiplex modeling should pay close attention to not only the inter-layer connectivity but also to the topologies of the layers, as they significantly impact diffusion of social information.

Taking these results together shows us several important aspects of multiplex diffusion dynamics using our operators.  First, the synthetic networks we tested show Fiedler values on the high side of the theoretical bounds we initially obtained, showing that we might expect on average faster diffusion in empirical networks.  Second, from the analysis of the twelve layer multiplex networks, we see that layer topology interacts with the rate of diffusion in a subtle way.  While the multiplex framework allows transfer along other layers to overcome slower diffusion in another layer, as the latter layer gains importance, the gain in the rate of diffusion decreases.  For the distinguished layers with the best connectivity --- for example, the Watts-Strogatz example ---  there is an optimal weight where the diffusion is fastest.  This last point is most interesting, indicating the interaction between the layer topologies is more than simply the sum of its parts.  We expect that multiple layers with diverse topologies will potentially exhibit even more complex behaviors.

 All of these results again point to the way the more general multiplex diffusion operator interpolates between the equi-distribution model and the reduction to a single-layer.  These results demonstrate an extreme example of super-diffusion where the multiplex diffusion is faster than in any of the layers considered alone.  In our case, one layer is promoted as $w$ tends to one while when $w=0$, the Fiedler value is identical to that of the aggregate network by Proposition \ref{prop:spec}.  The peak of the Fiedler value near $w=0.2$ for the model with the Watts-Strogatz layer is larger than the Fiedler values of any of the layers taken individually by a factor of $5$.  This indicates a new mechanism by which multiplex networks can exhibit super-diffusion.

The supra-Laplacian can exhibit super-diffusion as well, when the cross-layer diffusion rate is sufficiently large.  In that case,  the Fiedler value converges to the Fiedler value of the average aggregate network as the diffusion constant tends to infinity (see Figure 2 of \cite{gomez_diffusion_2013}).

\subsection{Stochastic Dynamics}\label{sec:rw}
Stochastic dynamics model probabilistic flows on networks. A left stochastic matrix is a non--negative matrix where the entries in each column sum to one.  Such a matrix defines a discrete Markov process on a network, with the entries $D_{ij}$ corresponding to the probability of moving from from node $j$ to node $i$ at each time step. Stochastic processes on networks are commonly used for determining centrality and detecting communities.

If a stochastic operator is primitive or irreducible, the Perron--Fr\"obenius Theorem guarantees that the largest eigenvalue of that matrix is $1$ and that the entries in one of the corresponding eigenvectors are non--negative. This vector represents the steady state of the dynamical system for arbitrary input.  When the Markov chain represents a random walk, we can interpret the entries as occupation probabilities.

If we use a closed general mixing model with layer operators given by stochastic operators $D^\alpha$, we have already noted that $\mathfrak{D}$ is also stochastic.  We'd also like to understand when $\mathfrak{D}$ is irreducible, which we approach by considering the equivalent network property of strong connectivity.  While this depends on the properties of the layer operators, it also depends on the connectivity of the matrix $C$.  If all the layers are strongly connected, then the strong connectivity of the single-layer network associated to $\mathfrak{D}$ depends on how easily we can move between the layers.  It is sufficient, for example, if when we aggregate each layer to a single node, the resulting network is also strongly connected.  To formalize this, we define a reduced inter-layer connectivity network:
\[\bar{C} = \begin{pmatrix} 1^T C^{1,1} 1 & \dots & 1^T C^{1,k} 1\\ \vdots & \ddots & \vdots \\ 1^T C^{k,1} 1 & \dots & 1^T C^{1,k} 1 \end{pmatrix} = \bar{I}^T C \bar{I}, \]
where $\bar{I}$ is the $nk \times n$ block diagonal matrix with blocks equal to the $n \times 1$ vector, $1$, consisting entirely of ones.  If $\bar{C}$ and all the layers are strongly connected, then $\mathfrak{D}$ is strongly connected as well.  To show this, we need to exhibit a path between the elements of any ordered pair $(n^\alpha_i,n^\beta_j)$.  Since $\bar{C}$ is strongly connected, there is a path from layer $\alpha$ to layer $\beta$ in the reduced network, given by the sequence $(\alpha=\gamma_1, \gamma_2, \dots, \gamma_\ell=\beta)$.  Consequently, $\bar{C}_{\gamma_m,\gamma_{m+1}}\neq 0$ and there exists $i_m$ with $c^{\gamma_{m+1},\gamma_m}_{i_m} \neq 0$ for each $m \in{1,\dots,\ell-1}$.  Using the strong connectivity of the layers, we construct paths within the layers between different nodes and then, via these non-zero inter-layer connectivity constants, can move across layers, eventually reaching the desired target.  With this argument, we've proved the following result.

\begin{proposition}\label{prop:stoch}
If each $D^\alpha$ is stochastic and $\mathfrak{D}$ is closed then $\mathfrak{D}$ is stochastic. Additionally, if all the $D^\alpha$ and $\bar{C}$ are irreducible then $\mathfrak{D}$ is irreducible.
\end{proposition}

We view a stochastic $\mathfrak{D}$ as describing a random walk on the multiplex network with the probability of transitioning from the copy of node $i$ on layer $\beta$ to the $\alpha$ layer copy of any node $j$ adjacent to $n^\beta_i$ is $c^{\alpha,\beta}_j \frac{1}{\operatorname{deg}(n^\beta_i)}$.

We can use the results of Proposition \ref{prop:spec} and the previous proposition to describe the steady-state vector.
\begin{corollary}\label{cor:stoch}  In each case below, we assume $\bar{C}$ is irreducible, that each $D^\alpha$ is stochastic and irreducible, and the hypotheses of Proposition \ref{prop:spec} hold.  Then,
\begin{enumerate}
\item $\mathfrak{D}_u$ and $D_a=D^1C^1+\dots +D^kC^k$ are stochastic.  If $w$ is the steady-state vector of $D_a$, then $v=(C^1w,\dots,C^kw)$ is the steady-state vector of $\mathfrak{D}_u$.
\item $\mathfrak{D}_h$ and $D_a=c^1D^1+\dots +c^kD^k$ are stochastic.  If $w$ is the steady-state vector of $D_a$, then $v=(c^1w,\dots,c^kw)$ is the steady-state vector of $\mathfrak{D}_h$.
\item $\mathfrak{D}_e$ and $D_a=\frac{1}{k}(D^1+\dots+D^k)$ are stochastic. If $w$ is the steady-state vector of $D_a$, then $v=\frac{1}{k}(w,\dots, w)$ where $w$ is the steady-state vector of $\mathfrak{D}_e$.
    \end{enumerate}
\end{corollary}

This Markov formulation is equivalent to the problem considered in \cite{trpevski_discrete-time_2014} for modeling distributed consensus which provides additional confirmation of our methods. Under similar hypotheses to those in \cite{domenico_navigability_2014,trpevski_discrete-time_2014}, our operator reduces to instances of their operators.  For example, using the generalized ranked layers model in this setting produces a stochastic matrix which is the supra-transition matrix in \cite{trpevski_discrete-time_2014} while the most flexible version of the normalized supra-Laplacian associated to the ``physical random walkers" in \cite{domenico_navigability_2014} coincides with the stochastic general mixing model. In the latter paper, we note that the operator the authors name the normalized supra-Laplacian is {\em not} a re-scaled version of the supra-Laplacian operator associated to the matched sum, defined in \cite{gomez_diffusion_2013} and described above.

\subsubsection{Random Walk Convergence} The magnitude of the second largest eigenvalue of the transition matrix associated to the random walk governs the rate of convergence of the random walk. We interpret the spreading process modeled by this operator as a discrete diffusion process. This approach was used to study transportation flow problems on the London subway and information or opinion spreading in social networks \cite{domenico_navigability_2014,trpevski_discrete-time_2014}. In these interpretations, the larger the second largest eigenvalue, the slower the information diffuses across the network.

Given their different structures, we expect the different multiplex formulations behave very differently by this measure. As the number of layers increases, aggregate networks generally become denser leading to a rapid convergence to the steady state. For the matched sum, increasing the number of layers generally leads to a decrease in overall density, as the only inter--layer edges are between copies of the same node, placing each node copy in a $k$--clique. This global sparsity and local density leads a random--walker to spend most of their steps moving between copies of the same node, which will slow the convergence to a steady state. Our stochastic operator $\mathfrak{D}$ is different than either of these, as a walker cannot move between copies of the same node and the global density is determined by the individual layers.  Consequently, we conjecture that in the limit as the number of layers tends to infinity, the second largest eigenvalue of the random walk for the aggregate model tends to zero, while the that of the random walk operator formed by normalizing matched sum models tend to one.  The eigenvalue for our models should lie somewhere in between.  

\begin{figure}[!h]
\centering
\includegraphics[height=2in]{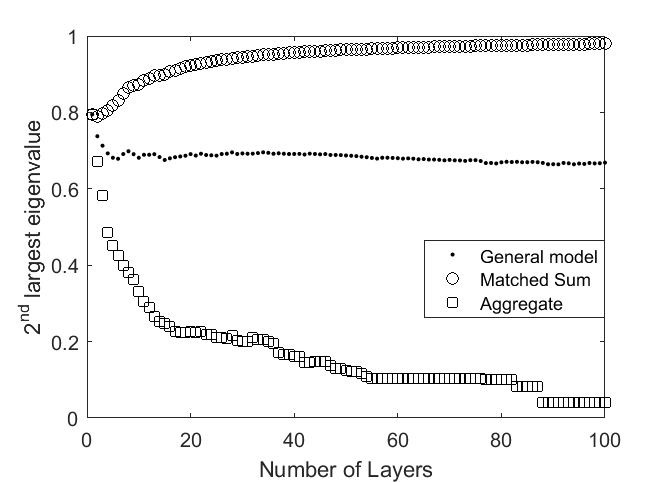}
\caption{Convergence rates of random walks on multiplex structures as $k\rightarrow\infty$. The behavior exhibited by our multiplex operators sits between the extremes of rapid diffusion (aggregation) and slow diffusion (matched sum). }\label{fig:rw}
\end{figure}

To evaluate these differences, we study the behavior of the random walk as the number of layers increases for a fixed number of nodes via simulation. We first generate a collection of $100$ Watts--Strogatz layers on 25 nodes.  For each of the aggregate, matched sum, and $\mathfrak{D}$ models, we construct $100$ increasingly complex models, where we construct the $\ell^\text{th}$ model in the list from the first $\ell$ layers from the collection.  For each of these, we build the associated random walk operator and compute the second largest eigenvalue. For the general mixing model, we choose the values of inter-layer constants proportional to the ratio of neighbors, $c^{\alpha,\beta}_i \propto \frac{\#\{n^\beta_j|n^\beta_j \sim n^\beta_i\}}{\#\{n^\alpha_j|n^\alpha_j \sim n^\alpha_i\}}$ and make the model closed. Figure \ref{fig:rw} shows the results, where we see evidence for our conjecture.  The results for other simplifications are similar.

As in our other applications, these results demonstrate the consequences of the differences between the models we examine.  They show an aspect of how our multiplex models interpolate between the aggregate and matched sum models.  Using other generative models for the layers and mixing coefficients again produces similar results. 

\subsubsection{Eigenvector Centrality}\label{sec:evcent}
A node has high eigenvector centrality if it is neighbors with other central nodes.  As with many measures of centrality \cite{borgatti_centrality_2005}, we can think about centrality using a dynamic process where we allow centrality values to flow between neighboring nodes until it reaches equilibrium.  In the single-layer case, the linear algebraic formulation of this notion yields the centrality as the values of the eigenvector of the adjacency matrix associated to its largest eigenvector.  As in other work in the multiplex literature \cite{de_domenico_ranking_2015,sola_eigenvector_2013}, we generalize eigenvector centrality to the multiplex setting using our framework in a straight-forward way --- by computing the eigenvector associated to the lead eigenvalue of our multiplex operator.  For some of the special cases of the operator $\mathfrak{D}$ when we take the $D_i$ to be the adjacency matrices of the layers, we recover existing results. For example, {\em uniform eigenvector-like centrality} and {\em global heterogeneous eigenvector centrality} \cite{sola_eigenvector_2013} are the lead eigenvector of the equi-distribution operator, $\mathfrak{D}_e$, and of the ranked layers operator, $\mathfrak{D}_h$, respectively.

We note that the eigenvector centrality in even the simplest version of our operator, the equi-distribution operator, cannot arise as a linear combination of the individual layer rankings. This is not a surprising result, since by Corollary \ref{cor:stoch} the spectral structure is related to determining the spectrum of a sum of matrices.  To see this precisely, it is easy to construct a simple multiplex network where this is true.

Using the full flexibility of our framework allows us to encode the importance of each layer to each node with a high level of specificity. We construct a three layer multiplex on 100 nodes with the layers created by the Barabasi--Albert preferential attachment process  with different intra-layer connection parameter for each layer. We construct different multiplex networks from this data using the equi-distribution, ranked layers, unified node, and general mixing approaches, as well as the matched sum and aggregate network for comparison. For the ranked layers model we choose the weights proportional to the density of each layer, for the unified node model we chose the weights to be proportional to the degree of each node copy, and for the general model we take the coefficients to be proportional to the number of common neighbors between copies.

\begin{figure}
\centering
\includegraphics[height=0.9in]{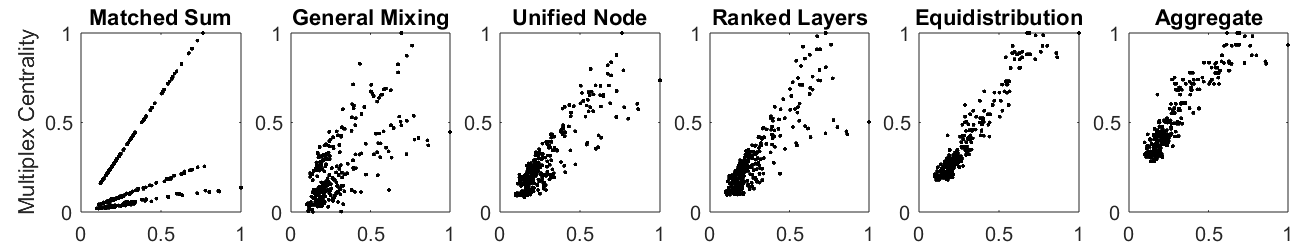}
\caption{Comparison of multiplex eigenvector centrality scores. The x--axis in each graph are the eigenvector centrality scores for the separate layer rankings for each node copy, while the y--axis reflects the multiplex centrality score. Varying the weighting scheme allows us to control how much mixing of centrality occurs between layers, while the matched sum model is just a linear transformation of the original rankings. }\label{fig:ev}
\end{figure}

For each of these multiplex representations, we compute the eigenvector centrality by finding the eigenvector associated to the largest eigenvalue of the associated operator, normalized so the values are between $0$ and $1$. We compare these centrality scores to baseline centrality scores --- the centralities of the nodes in the disjoint layers model.  In Figure \ref{fig:ev}, we plot the centralities of the multiplex models against the layer centralities to visualize the differences between the formulations.  A point $(x_{i,\alpha},y_{i,\alpha})$ in these graphs shows the centrality of node $i$ in the single-layer network given by layer $\alpha$ ($x_{i,\alpha}$) and the multiplex centrality of $n_i^\alpha$ ($y_{i,\alpha}$).  For the aggregate network, the centrality of $n_i^\alpha$ is the same for all $\alpha$ --- it is the centrality of node $i$ in the aggregate network.

The Figure shows us a clear interpolation between the aggregate and matched sum formulations.  The leftmost graph shows that the centralities of the matched sum are the same as the disjoint layer centralities up to a collection of linear transformations.  Moving through other formulations, we see the layer centralities combined in more and more complex ways until we reach the centralities of the aggregate network.  Consequently, the different weighting formulations allow us to sample from a continuum of results for a given multiplex data set, where the choice of weighting should be tailored to the particular data. Repeating these experiments for other mixtures of generative models including Erd\"os--R\'enyi and Watts--Strogatz yields similar results.

\section{Applications}\label{sec:apps}
We return to our two empirical examples to examine them through the lens of our multiplex dynamics framework.  Our motivation is two-fold.  First, we wish to get a sense of how ``real" networks behave compared to the theoretical results as well as those derived from simulating synthetic networks.  Second, both of these empirical networks provide important models in the social sciences and the application of our framework provides a new avenue for analysis with the potential to reveal new features of the underlying systems.
\subsection{The World Trade Web}\label{sec:appwtw}
The World Trade Web is frequently studied as an economic complex network. Original analyses of the data were carried out on an aggregate structure \cite{Fagiolo_2008,Fagiolo_2010}, but more recent research has reexamined this object from a multiplex perspective, discovering that the properties of the multiplex are quite different from the aggregate with regards to measures such as reciprocity, assortativity, clustering coefficients, and other standard network metrics\cite{Barigozzi_2010,Barigozzi_2011}. Here we analyze data from the 2000 World Trade Web \cite{Feenstra2005}, disaggregated into the 10 commodity layers by the United Nations' Standard International Trade Classification (SITC) one-digit code, listed in Table 2 below.  The heterogeneity in intra-layer activity as well as transitivity indicates that we should expect observed differences between the network properties of the aggregated network and the multiplex formulation.

In Section \ref{sec:motivation}, we outlined why the WTW is a good candidate for our multiplex model. The layer dynamics model the flow of dollars between countries for a particular commodity type. The unified node model extends these commodity transactions to the whole structure by aggregating the total incoming dollars from all layers, and then redistributing them according to the needs of each layer. As discussed in Section \ref{sec:uninodes}, we use the unified node model with the $c^\alpha_i$ given by the proportion of the in-degree on layer $\alpha$ and the total in-degree.

\begin{table}[h]
\centering

\begin{tabular}{ccccc}
\hline
Layer &Description&Volume&\% Total&Transitivity\\
\hline
0& Food and live animals&291554437&5.1&0.82\\
1& Beverages and tobacco&48046852&0.9&0.67\\
2& Crude materials&188946835&3.3&0.79\\
3& Mineral fuels&565811660&10.0&0.62\\
4& Animal and vegetable oils&14578671&0.3&0.64\\
5& Chemicals&535703156&9.5&0.83\\
6& Manufactured Goods&790582194&13.9&0.87\\
7& Machinery&2387828874&42.1&0.85\\
8& Miscellaneous manufacturing&736642890&13.0&0.83\\
9& Other commodities&107685024&1.9&0.56\\
\hline
All&Aggregate Trade& 	5667380593&100&0.93\\
\hline
\end{tabular}
\caption{Layer information for the 2000 World Trade Web, showing the $10$ layers of the top level disaggregation of the trade data, their overall trade volumes, their percentage of the aggregate trade, and the transitivity of the layer network.}
\end{table}

 As in Section \ref{sec:evcent}, we use a centrality measure to compare the disjoint layer rankings to the multiplex and aggregate rankings, but instead of eigenvector centrality, we use random walk betweenness centrality (RWBC) \cite{newman2005}.  A node $i$ has high RWBC if a random path between two arbitrary nodes is likely to pass through node $i$.  Another way to view this centrality in the WTW networks is release an additional dollar into the network at a random node and watch its propagation through network according to trade exchanges selected at random from those available at each step.  Nodes that are likely to have this dollar pass through them will have higher random walk betweenness centrality.  For our multiplex model, we use the operator $\mathfrak{D}$ with the intra-layer dynamics given by random walk operators and the inter-layer dynamic parameters given by Equation \eqref{WTW_params}.

 Another version of RWBC for multiplex networks, based on the matched sum, is described in \cite{sole-ribalta_random_2016}. The model presented here differs in two fundamental ways. First, the previous version allows for the random walker to transition directly between copies of the same node to move between layers, while our model as described in \eqref{eqn:la_unified} does not, following the discussions of Section \ref{sec:rw}. Secondly, the aim of the method in the previous paper was to determine a single ranking for each node by averaging over the layers, whereas our method attempts to assign a separate centrality score to each node copy. Thus, the two RWBC measures are capturing distinct information about the network, even though they employ the same underlying concept.  

 We compare the RWBC of the unified node model to those of the disjoint layers and average aggregate models, computing the rank correlations between their centrality scores.  All correlations are relatively large, over $0.5$, but there are numerous significant differences between the rankings for specific nodes.  Table \ref{tab:bigdiffs} shows the countries and commodity layers where the orderings between the unified node RWBC and the aggregate RWBC are the largest over all layers.  In parentheses, we show the difference between the unified node rankings and the disjoint layer rankings for comparison.  To interpret the entries in the table consider, for example, Saudi Arabia's layer $3$.  The entry in the top table is $-20(-3)$ which indicated that the unified node ranking for layer 3 is $20$ places down than that of the aggregate but only $3$ places down from that of the disjoint layer network.

\setlength{\tabcolsep}{4pt}
\begin{table}
\scriptsize
\begin{center}
\begin{tabular}{c|cccccccccc}
\hline
 & 0 & 1 & 2 & 3 & 4 & 5 & 6 & 7 & 8 & 9  \\
\hline
Saudi Arabia & +37 (-1) &+31 (-2)& +32 (+7)& -20 (-3)& +32 (-1)& +8 (+5)& +31 (-2)& +27 (+3)& +31 (+3)& +25 (+2) \\
Iran & +9 (+4) &+16 (-2)& +13 (+3)& -24 (-2)& +20 (+0)& +15 (+2)& +14 (+3)& +20 (+3)& +20 (+2)& +22 (+9) \\
Kazakhstan & -17 (-20) &-10 (-17)& -37 (-10)& -39 (+11)& +5 (-5)& -13 (-11)& -24 (-12)& -7 (-10)& +5 (-4)& -14 (-9) \\
Kuwait & +19 (+0) &+18 (-1)& +18 (+4)& -24 (-3)& +15 (-1)& +10 (+1)& +19 (+0)& +16 (+3)& +16 (+0)& +14 (+3) \\
Algeria & +15 (+0) &+11 (+7)& +14 (+1)& -29 (-3)& +15 (+3)& +11 (+6)& +15 (+1)& +15 (+3)& +15 (+1)& -17 (+0) \\
Nigeria & +13 (+7) &+17 (+5)& +12 (+4)& -25 (-1)& +8 (+8)& +17 (+4)& +15 (+2)& +17 (+1)& +17 (+0)& +15 (+2) \\
Peru & -27 (+6) &+1 (+9)& -29 (+8)& -13 (-9)& -34 (-2)& -5 (-4)& -12 (-3)& +1 (-4)& -8 (-2)& -22 (-8) \\
China & +22 (+0) &+4 (-11)& +20 (-1)& +31 (+1)& +21 (+15)& +9 (-8)& -4 (-10)& -5 (-2)& -18 (+0)& -7 (-6) \\
Chile & -25 (+1) &-27 (+0)& -32 (+2)& +17 (+0)& -6 (-1)& -1 (-1)& -9 (-1)& +9 (+0)& +12 (-1)& -1 (+0) \\
Japan & +33 (+7) &+24 (-5)& +17 (+3)& +29 (+5)& +23 (+11)& +2 (-5)& +0 (-4)& -1 (+0)& +2 (+2)& +3 (+0) \\
\hline
\end{tabular}
\end{center}
\caption{Countries with large differences between layer rankings of the unified node RWBC and the RWBC of the aggregate network.  We show the differences with the disjoint layer rankings in parentheses for comparison.}\label{tab:bigdiffs}
\end{table}
\setlength{\tabcolsep}{6pt}

We see a particularly interesting pattern in the third layer, Mineral Fuels, which includes petroleum products, where we see the top countries have significant differences between the rankings, with the multiplex ranking substantially higher than the aggregate ranking, but relatively similar to the disjoint layer rankings.  This demonstrates that in some cases, countries like Saudi Arabia, Iran, Kuwait, Algeria and Nigeria, with an emphasis on one layer due to substantial oil resources, have that importance diluted in the aggregate network but still highlighted in the multiplex network.  All of these countries have trade asymmetry, with large amount of trade in the third layer but much less in others, contributing further to this effect.  Countries with more uniform trade distributions, such as the United States, still have large volumes of trade in the third layer, but do not see large mismatches in rankings between the unified node and the aggregate or disjoint layer rankings.

These analyses indicate that for RWBC, the unified node model is a coarse interpolation between the aggregate and disjoint layer models, although not as transparently as we saw with eigenvector centrality.  While the rank correlations between the centralities do not support a dependence between the two, looking at the countries with large deviations shows that in those cases the unified node rankings are closer to the disjoint layer rankings that the aggregate ones.  However, there are many examples --- Kazakhstan is one seen in our table above --- where the rankings are all very different from one another.  This is likely due to differences between RWBC and eigenvector centrality.  The latter depends only on the rank one approximation of the operator and hence, throws away a lot of subtle dynamical information.  On the other hand, RWBC depends on the long term behavior of the random walk process and therefore incorporates more varied aspects of the dynamic process.
\subsection{Information flow in a social multiplex network}\label{sec:social}
 As we described in Section \ref{sec:introsocial}, to study network effects on the flow of information about microfinance opportunities, researchers surveyed $75$ villages in the Karnakata region in India about their social connections across twelve categories \cite{Banerjee2013}.  Table \ref{tab:villages} describes the individual layers and the density of each layer, as well as the number of connected components and proportion of the layer nodes contained in the giant component for the two villages we'll use in our application.  Looking at the multiplex network shows that all of its layers are very sparse and have many connected components.  Consequently, analysis of the layers individually misses a fundamental property of the multiplex network.  On the other hand, the aggregate network has many fewer components, with the giant component containing almost all the nodes.  Again, we see our multiplex model lying between these two extremes.

 \begin{table}
\scriptsize
\centering
\begin{tabular}{c|c|ccc|ccc}
\hline
& & \multicolumn{3}{c|}{Village 5}&\multicolumn{3}{c}{Village 61}\\
\hline
Layer&Description&Density& Components&Giant \% &Density& Components&Giant \% \\
\hline
\hline
0& Borrow Money&.0082&26&.8354&.0108&15&.9188\\

1&Give Advice&.0077&49&.5892&.0098&34&.7377\\

2&Help Make Decisions&.0076&61&.1277&.0100&24&.8562\\

3&Borrow Kerosene or Rice&.0085&21&.8338&.0113&14&.9171\\

4&Lend Kerosene or Rice&.0086&22&.8308&.0113&14&.9255\\

5&Lend Money&.0081&14&.7908&.0107&17&.9036\\

6&Medical Advice&.0075&84&.2938&.0106&14&.9306\\

7&Friends&.0089&15&.9277&.0105&22&.8714\\

8&Relatives&.0085&29&.7231&.0105&26&.5448\\
9&Attend Temple With&.0073&117&.0462&.0089&108&.0372\\

10& Visit Their Home&.0087&15&.9185&.0116&11&.9475\\

11& Visit Your Home&.0088&16&.9108&.0117&11&.9492\\
\hline
All&Aggregate&.0121&3&.9862&.0155&8&.9679\\
\hline
\end{tabular}
\caption{Layer information for two Karnakata villages with the density, number of connected components, and proportion of nodes contained in the giant component for each layer. }\label{tab:villages}
\end{table}

\begin{figure}
\centering

\includegraphics[height=3in]{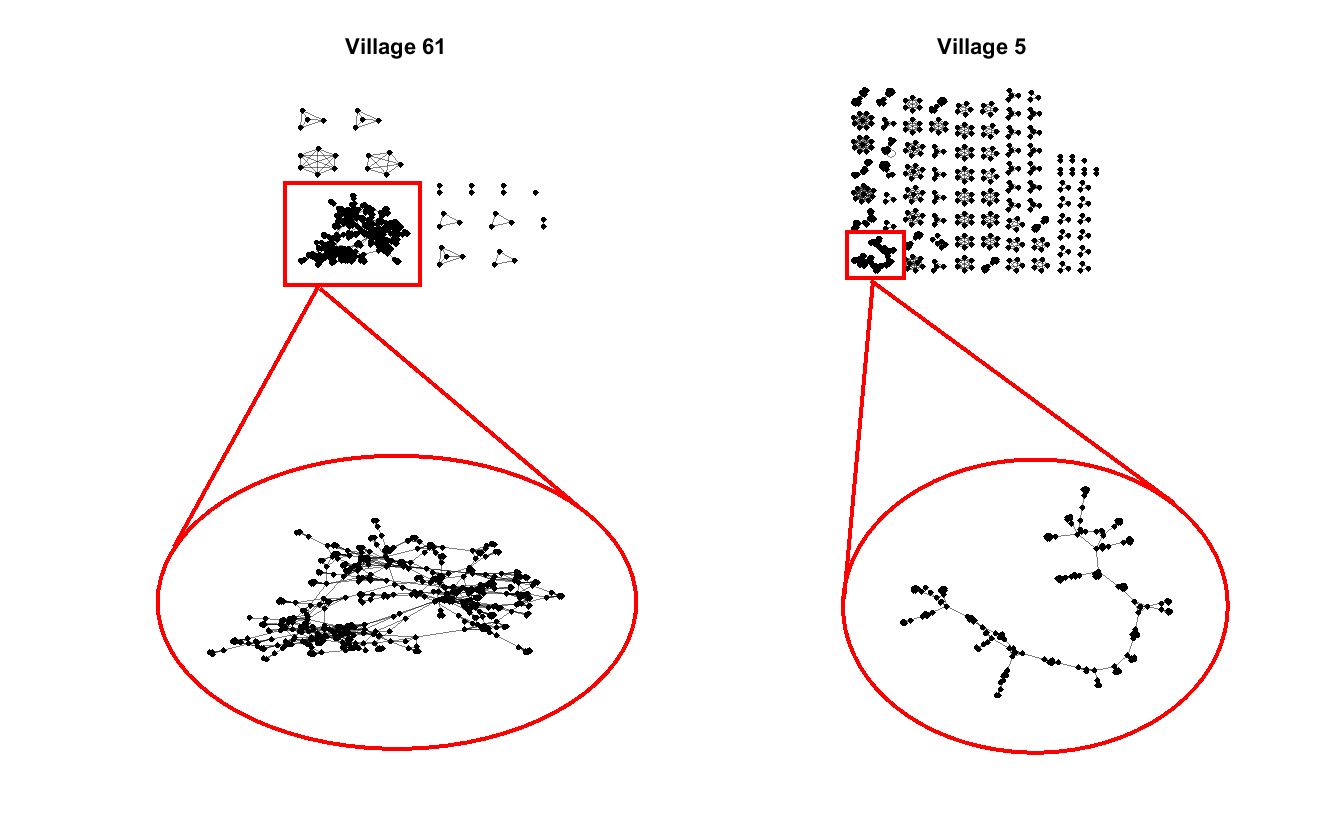}
\caption{Visualizations the giant component of the medical advice layer and the giant component of the medical advice layer for two of the Karnakata villages. In village 5 the medical layer consists almost entirely of small cliques, while in village 61 the medical layer has a large giant component.  }\label{fig:villtop}
\end{figure}

We will mimic the earlier analysis on synthetic networks \ref{sec:ratediffusion} to measure the effects of prominence the ``medical advice" layer when considering the diffusion of medical information across the social network. There are two distinct structural archetypes that appear in the medical advice layers across the villages, typified by the examples described in Table \ref{tab:villages} and shown in Figure \ref{fig:villtop}. In some of the villages, such as village 5, the medical advice layer consists almost entirely of small cliques with a small path--like connected component. Alternatively, villages like village 61 have a large connected component with several distinct clusters.   We will look at two node subsets for each village --- the giant component of the aggregate network and the giant component of the medical layer.  For the giant component of the medical network, Figure \ref{fig:villtop} indicates that we might expect Village 5 to behave similarly to the ``single path" topology of Section \ref{sec:ratediffusion} and Village 61 to behave like the small-world topology.  When we restrict to the aggregate's giant component, the medical layer is disconnected with Village 5 having many more connected components than Village 61. Table \ref{tab:sw_vill} shows us that under the restriction to the medical layer's giant component, both Villages exhibit small-world properties.

\begin{table}
\centering
\begin{tabular}{cccc}
Village & Type & Clustering Coefficient & Pathlength \\\hline
5 &Aggregate & 0.91 &  $\infty$ \\
5 &Medical &0.84  &  11.02 \\
 & Null &0.03  &  3.17 \\
 & Ratio &27.91 &   3.47 \\
 61 & Aggregate &0.76  &  $\infty$\\
 61 & Medical &0.76 &   6.50 \\
& Null &    0.01 &   3.57 \\
&Ratio & 66.03   & 1.82 \\ \hline
\end{tabular}
\caption{Small-world characteristics of Karnakata Village networks.  When restricting to the aggregate giant component, the medical layers are disconnected, making their average pathlengths infinite.  However, when restricting to the medical layer's giant component, layer networks for both villages fall into the small-world category.}\label{tab:sw_vill}
\end{table}

\begin{figure}
\centering
\includegraphics[height=2in]{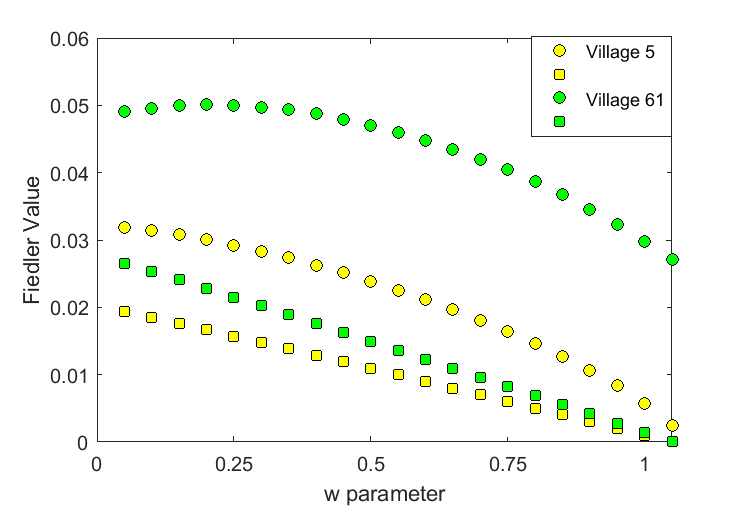}
\caption{Diffusion rates for medical information over the aggregate giant component (squares) and the medical layer giant component (circles) for Village 5 (yellow) and 61 (green) as the weight parameter varies from $1/12$, the equi-distribution model, to $1$, where all weight is placed on the medical layer alone.}\label{fig:fiedler_villages}
\end{figure}

As we saw in Section \ref{sec:diffusion}, these types of connection topologies have a strong impact on the flow of information within the multiplex.  Using the coefficients for the ranked layers model describe in Section \ref{sec:hierlay}, we varying $w$ on $[0,1]$ to examine the effect of the importance of the medical advice layer.   For our layer dynamics, we'll again use the graph Laplacians associated to the layers as a simple model of information diffusion.   As, even in the multiplex setting, the networks have multiple connected components, we'll consider restriction to the giant component of the aggregate network as well as to the giant component of the medical layer.  Figure \ref{fig:fiedler_villages} summarizes the Fiedler values as a function of $w$.  We next describe the results in detail.

In the first case, although we have restricted to the aggregate giant component, the medical advice layers are still disconnected for both villages. Thus, as $w$ tends to one the Fiedler value goes to zero for both operators.  Again, we see that the selection of weights in our operator allows us to interpolate between extreme values of the aggregate, represented by the equi-distribution value at $w=\frac{1}{12}$, and the disconnected medical advice layer. The Fiedler values for these systems are nearly linear as a function of $w$ with village 5 fitting $-0.0212w+0.021$ and Village 61 fitting $-0.02901w+0.02924$ both to within $10^{-6}$ sum of squares error.  This is consistent with the result for the synthetic networks in Section \ref{sec:ratediffusion} as both villages have topologies similar to the disjoint triangles topology.

In the second case, where we restrict to the giant component of the medical layer, we observe some differentiation between the behavior of the Laplacian operators as $w$ varies. Here we see that for village 61, with the dense medical advice layer, the diffusion rate improves for values of $w$ less than roughly $0.25$ before peaking and decreasing towards zero.  For village 5 we also observe a piece of a concave down curve, but with no local maximum. 	We note that the village with the better connected medical layer, Village 61, has a higher Fiedler value for each $w$. However, these values are not directly comparable as the villages have different populations and underlying density.  Our results for synthetic networks in Section \ref{sec:ratediffusion} place these results in context. As hypothesized, both villages, albeit to different extents, behave like the synthetic multiplex network with the Watts-Strogatz medical layer.

These results confirm our conclusions from the work on synthetic networks in an empirically derived layer networks.  We see that with respect to the diffusion of medical information, our model suggests there is an optimal weighting trading off between the medical layer and others which facilitates the fastest spread of information.  This is a particularly stark example of the super-diffusion discussed in Section \ref{sec:ratediffusion} as, individually, the layers are not connected and hence the time scale of complete diffusion is infinite for the disjoint layers model.  Even when restricting to the giant component of the medical layer, many of the other layers still have multiple components.

\section{Conclusion}

We've introduced a flexible framework for modeling dynamics on multiplex networks for which treating node copies as distinct entities is inappropriate.  Our framework avoids direct interaction between node copies, a feature inherent in many structural formulations.  For stochastic processes, our operator matches or generalizes others in the literature, while our model of diffusion is distinct from others in the literature.  In both cases, we apply spectral results for broad classes of models within our framework to illuminate properties of the diffusion and stochastic multiplex processes.  In particular, our framework provides a new way to understand the interaction between layer processes which results in an overall process which is more efficient than the layer processes treated separately.

This last point ties into a broader observation: operators within our framework interpolate between two simple models of multiplex systems -- the average aggregate network and the disjoint layers model.  This flexibility is particularly useful in investigations of networks, such as the empirical trade and social networks we study in Section \ref{sec:apps}, where we can tune the properties of our model to fit properties of or constraints on the system in question.

Our family of operators provides a platform for extending dynamically motivated network analyses effectively to the multiplex network setting for the systems they model well.  We demonstrate an example of this in defining a multiplex eigenvector centrality, which extends some other existing multiplex centralities.  Future work includes extending dynamical methods of community detection.

\bibliography{Multiplex}

\end{document}